\documentclass[aps,twocolumn,showlabels,showrefs,amsmath,amssymb,pre,superscriptaddress,floatfix,colors,longbibliography]{revtex4-1}

\usepackage{graphicx}% Include figure files
\usepackage{dcolumn}% Align table columns on decimal point
\usepackage{bm}% bold math
\usepackage{amssymb}
\usepackage{multirow}
\usepackage{color}
\usepackage{hyperref}
\usepackage[normalem]{ulem}
\usepackage{float}
\usepackage{mathrsfs}
\usepackage{soul}
\usepackage[T1]{fontenc}

\begin{document}

\title{Ion channels in critical membranes: clustering, cooperativity, and memory effects}

%\date{\today}

%\pacs{..}

\author{Antonio Suma} 
\affiliation{Dipartimento di Fisica, Universit\`a degli Studi di Bari and INFN, Sezione di Bari, via Amendola 173, Bari, I-70126, Italy}
\affiliation{Institute for Computational Molecular Science, Temple University, Philadelphia, PA 19122, USA }
\author{Daniel Sigg} 
\affiliation{Institute for Computational Molecular Science, Temple University, Philadelphia, PA 19122, USA }
\affiliation{dPET, Spokane, WA,  USA}
%\author{Abhinav Srivastava}
%\affiliation{Institute for Computational Molecular Science, Temple University, Philadelphia, PA 19122, USA }
%\author{Jason Pattis}
%\affiliation{Institute for Computational Molecular Science, Temple University, Philadelphia, PA 19122, USA }
%\author{Lucia Coronel}
%\affiliation{Institute for Computational Molecular Science, Temple University, Philadelphia, PA 19122, USA }
\author{Seamus Gallagher}
\affiliation{Institute for Computational Molecular Science, Temple University, Philadelphia, PA 19122, USA }
%\author{Vincent Voelz} 
%\affiliation{Institute for Computational Molecular Science, Temple University, Philadelphia, PA 19122, USA }
\author{Giuseppe Gonnella} 
\affiliation{Dipartimento di Fisica, Universit\`a degli Studi di Bari and INFN, Sezione di Bari, via Amendola 173, Bari, I-70126, Italy}
\author{Vincenzo Carnevale} 
\affiliation{Institute for Computational Molecular Science, Temple University, Philadelphia, PA 19122, USA }

\begin{abstract}

%Much progress has been made in elucidating the internal mechanism for voltage-dependent gating in ion channels, but less well-understood is the influence that lipid raft formation in a fluid mosaic membrane has on gating kinetics. Here we propose that state-dependent channel affinity for different lipid species contributes toward experimental behaviors of channel clustering, cooperativity, and hysteresis. We employ models of voltage-gated ion channels engaged in Ising-like interactions with membrane lipids close to the critical temperature to demonstrate the potentially universal basis for these anomalous behaviors.

Much progress has been made in elucidating the inner workings of voltage-gated ion channels, but less
understood is the influence of lipid rafts on gating kinetics.
Here we propose that state-dependent channel affinity for different lipid species provides a unified explanation for the experimentally observed behaviors of clustering, cooperativity, and hysteresis.  We develop models of diffusing lipids and channels engaged in Ising-like interactions to investigate the collective behaviors driven by raft formation in critical membranes close to the demixing transition. The model channels
demonstrate lipid-mediated long-range interactions,  activation curve steepening, and long-term memory in ionic currents.
These behaviors likely play a role in channel-mediated cellular signaling and suggest a universal mechanism for self-organization of biomolecular assemblies.

\end{abstract}

\maketitle

\noindent

Ion channels play a critical and ubiquitous role in cellular signaling by transmuting external forces into changes in ion permeability across membranes, a process known as gating. The discovery of on/off ionic currents catalyzed by a single pore~\cite{katz1972statistical,anderson1973voltage} led to a reinterpretation of the seminal Hodgkin and Huxley equations~\cite{hodgkin1952quantitative} 
as stochastic gating by independent channels~\cite{colquhoun1995principles}.
%as emergent population-level descriptions of collections of isolated proteins undergoing stochastic gating. 
As electrophysiological and biochemical techniques improved,  gating schemes expanded to accommodate new data ~\cite{Sigworth1994Review, Bezanilla2018Review}, but the assumption that channels gate independently has persisted. 
%(however, see Dilger, 1993).
 
The paradigm of the independent channel has been challenged by experiments that directly visualize the cellular membrane, such as electron microscopy, confocal fluorescence microscopy, and superresolution imaging~\cite{shelley2010coupling,sato2019stochastic,pfeiffer2020clusters,vierra2021regulation}. Channels are seen to aggregate in clusters ranging in size from a few channels to large groups containing tens of thousands of channels. Factors that contribute to   clustering include direct channel-channel interactions~\cite{moreno2016ca2+} and connections to the cytoskeleton~\cite{Lillemeier2006Cytoskeleton}. 

Here we consider another mechanism of clustering that is mediated by the lipid membrane. The fluid-like dynamics of a heterogeneous population of lipids can indirectly mediate channel interactions. The notion that conformational changes in the channel can affect and be affected by the thermodynamic properties of the membrane is supported by a range of evidence from cryoEM, electrophysiology, and computational studies~\cite{liu1993application,seeger2010changes,machta2012critical,hite2014phosphatidic,katira2016pre,kimchi2018ion,duncan2020lipid,duncan2020defining,bodosa2020preferential,mandala2022voltage,levental2022regulation}. Lipid membranes under physiological conditions are close to the demixing transition, resulting in the formation of rafts and liquid disordered domains~\cite{veatch2003separation,sezgin2017mystery, shaw2021critical,pantelopulos2018regimes}. Near the critical temperature, membranes are subject to fluctuations in local composition characterized by large correlation lengths~\cite{honerkamp2009introduction}.

Membrane demixing can explain two observations that are currently poorly understood: (i) the unusual size distribution of clusters of ion channels, which can follow a power-law behavior, see Supplementary Material (SM)~\footnote{See Supplemental
    Material (SM) for the models details and additional analysis. The SM includes additional
Refs.~\cite{Coniglio2009,Clauset_2009,shinoda2004rapid,LAMMPS,kemmer2010nonlinear,Colquhoun1995Qmatrix,stefani1994gating,zagotta1994shaker,senning2015activity,Peters_rot_diff_1982,fernandez2023inactivation,leonard2021activation,Chowdhury2012VM,Chowdhury2010Hill,gillespie1977exact,Gillespie2007,Sigg1997,Sigg1999Kramers}}  analysis from Ref.~\cite{sato2019stochastic}, and (ii) the long-term memory effects seen in single channel recordings~\cite{VARANDA2000Hurst,Siwy2001HurstBK,Bandeira2008HurstBK,wawrzkiewicz2017temperature,Wawrzkiewicz2018HurstBK,silva2021memory}. Furthermore, since interactions with specific lipid species have been shown to selectively stabilize the open or closed state of the pore~\cite{faure2014lipids}, near-critical long-range lipid interactions offer explanations for other anomalous channel behaviors, such as enhanced cooperativity~\cite{moreno2016ca2+,dixon2022mechanisms}, modal gating~\cite{PATLAK1979Modal,Magleby1983Modal,siekmann2014statistical} and hysteresis~\cite{bezanilla2013gating,cowgill2023charge,villalba2020hysteretic}.

In this work, we propose a unified explanation for the size distribution of clusters and gating anomalies based on the assumption that  distinct conformational states of the channel (including open and closed) have affinities for different lipids species. To this end, we introduce two complementary phenomenological descriptions of ion channels embedded in a lipid mixture: i) a mesoscopic model of lipid membranes, which faithfully describes the membrane phase behavior and its coupling to the channel conformation; and ii) a discrete-state Markov model of interacting channel-lipid systems with a realistic description of channel gating kinetics. \textcolor{black}{ The first model is a minimalist representation of lipids and channels whose purpose is to demonstrate that these collective phenomena are independent of any specific molecular detail, but rather derive from the combination of state-dependent interactions and critical demixing. The second model builds on a more realistic representation of the channel structure and activation dynamics and allows us to ascertain that the effects investigated here occur on physiologically relevant length- and time-scales. In both models,} we find that fluctuations in the underlying lipid medium  give rise to long-range effective attractive forces leading to clustering, cooperative gating, hysteresis, and long-memory effects in the channel gating dynamics.

\section{results}

\subsection{A mesoscopic model of ion channels in phase-separating membranes}

To study the dynamics of channels embedded in a lipid bilayer, we introduce a simplified mesoscale representation of the membrane and of the embedded ion channels.

\begin{figure*}[h]
\centering %1.5cm 0.5cm 2.3cm 9cm
\includegraphics[width=1.9\columnwidth]{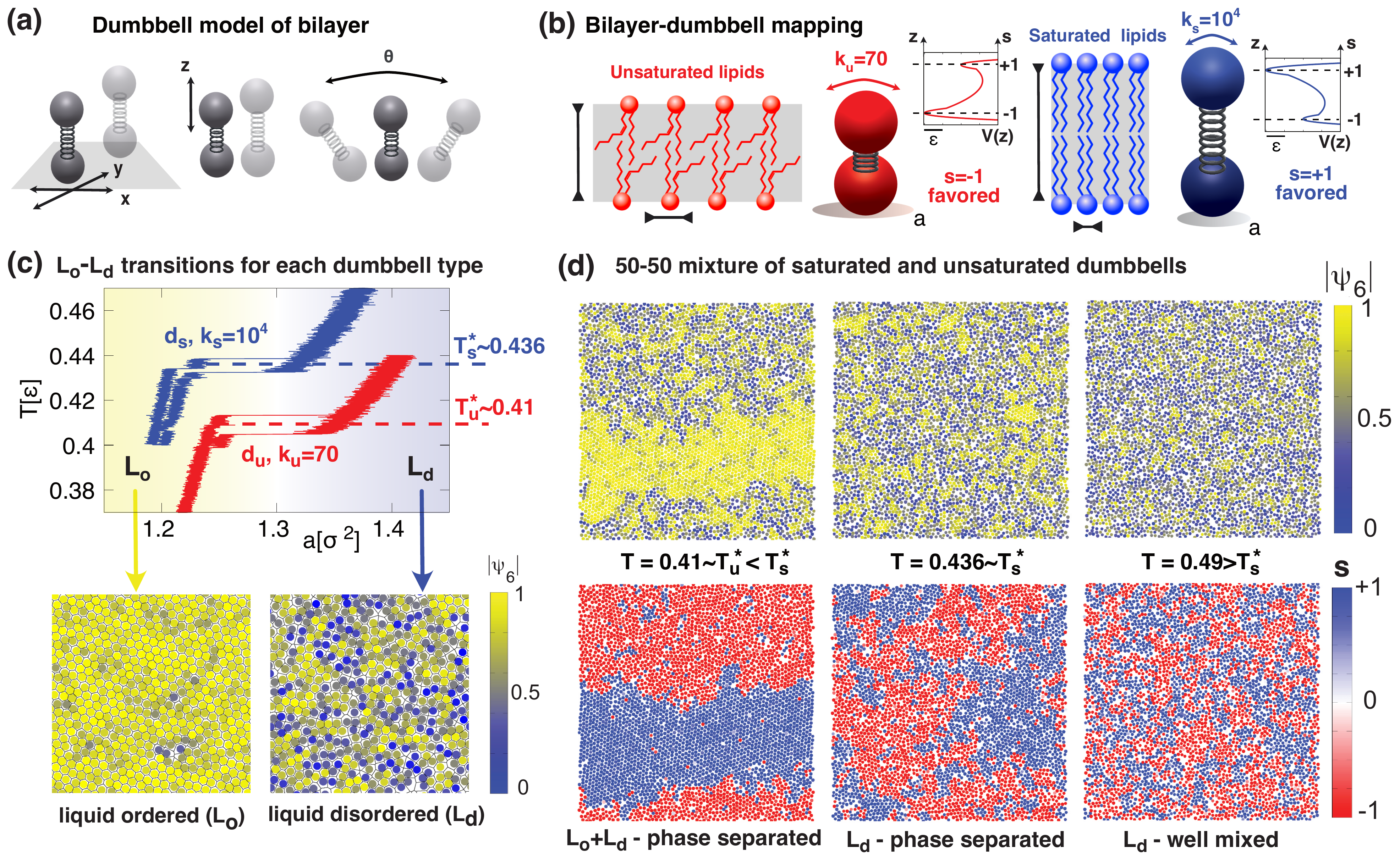}   \\
\caption{(a) Mesoscopic model of lipids, represented by a dumbbell which can diffuse  on the  xy-plane (bottom bead constrained in plane), has a varying z-height of top bead, and a deviation ($\theta$ angle) from the vertical orientation  restrained by a harmonic potential of force constant $k$ (SM). 
 (b) Bilayer-dumbbell mapping. Unsaturated (saturated) lipids are mapped onto $d_u$ ($d_s$) dumbbells, with a global minimum of the height potential (see $V(z)$) corresponding to short (long) dumbbell or spin s=-1 (+1), and a small (large) tilting constant, $k_u=70$  ($k_s=10^4$),  which entails a large (small) surface area  $a$.  (c) Plots of temperature versus $a$ for  $d_u$ and $d_s$, obtained via heating/cooling the system (two curves for each type). Both dumbbell types exhibit a liquid ordered ($L_o$, associated to hexatic ordering) to liquid disordered ($L_d$) phase transition, see snapshots colored with local hexatic modulus $|\psi_6|$ (SM). As curves do not coincide due to hysteresis~\cite{katira2016pre}(ordering from the  disordered phase is much slower than disordering from the ordered phase, we considered as transition temperature the average value in between the heating/cooling curves plateau, $T_u^*\sim0.41$ and $T_s^*\sim0.436$. The melting temperature is shifted towards higher $T$ as $k$ is increased.
 (d)  50-50 mixture of $d_u$ and $d_s$ dumbbells at three temperatures, $T=0.41, 0.436, 0.49$. Each column shows an equilibrated conformation for the indicated temperature, colored by the hexatic modulus and spin value (top and bottom row, respectively).}
\label{fig1}
\end{figure*}

The membrane is described by an assembly of molecules of two types, each representing saturated or unsaturated lipids. The assembly is meant to mimic the phase behavior of a three component membrane in physiological conditions~\cite{veatch2005seeing,garcia2007effect}. In cholesterol containing ternary mixtures with saturated and unsaturated lipids, \textcolor{black}{saturated lipids} tend to segregate into liquid ordered ($L_o$) domains, while \textcolor{black}{unsaturated lipids} promote the formation of a liquid disordered phase ($L_d$). $L_o$ domains are characterized by hexatic order (see \cite{katira2016pre,javanainen2017nanoscale,pantelopulos2018regimes,gu2020phase} and SM for a definition) and a larger membrane thickness compared to the $L_d$ phase, see Fig.~\ref{fig1}b. At fixed temperature, the transition from a well-mixed liquid disordered membrane to a phase-separated coexistence of $L_o$ and $L_d$ domains is driven by a change in lipid composition ({\it i.e.} a change in the cholesterol and lipids molar fractions). Importantly, at fixed composition, this mixing/demixing transition can be driven by a change in temperature: higher temperatures promote mixing of the two lipids into a liquid disordered ($L_d$) phase.

\textcolor{black}{We model each ion channel as a multi-domain inclusion in the membrane in which each domain is allowed to fluctuate across a single-barrier conformational landscape. The stable basins are meant to recapitulate the two-state character of ion channels that cycle through open/closed (in the case of the central pore unit) and resting/activated (in the case of ancillary sensory domain) conformations . We further assume that the open and closed or activated and resting conformations have affinity for saturated and unsaturated lipids, respectively.} This feature of the model is meant to mimic the dependence of ion channel activation on the properties of the embedding membrane: this has been observed for KvAP channels, which have been shown to respond to the membrane potential differently if immersed in a bilayer of saturated or unsaturated lipids \cite{faure2014lipids} and for the Eag Kv channel whose activation entails a disorder/order transition in the surrounding lipid membrane~\cite{mandala2022voltage}. These observations raise the intriguing possibility that local lipid composition ({\it i.e.} the ratio between saturated and unsaturated lipids) and degree of orientational order ({\it i.e.} $L_o$ vs $L_d$) can potentially affect the activation properties of channels.

We built a model for the lipid bilayer by extending the particle-based Ising model described in Ref.~\cite{novinger2021particle}, Fig.~\ref{fig1}a and SM. In the particle-based Ising, basic building blocks are connected pairs of beads ({\it i.e.} a dumbbell) orthogonally oriented with respect to the membrane plane. The bottom beads are constrained to lay on the plane, and interact only through a volume exclusion short-range repulsion, while the top beads interact via a Lennard-Jones potential ({\it i.e.} it also includes an attractive tail). The z position of the top bead can assume two values (the two minima of a quartic potential); we showed in Ref.~\cite{novinger2021particle} that these two top-bead heights can be mapped to the spin values -1 and +1, and the dumbbells, fixed in a triangular lattice, have a ferromagnetic/paramagnetic transition varying the temperature. This work extends the previous model by allowing  dumbbells to freely diffuse  (xy-directions), and the dumbbell vertical orientation (described by the tilting angle $\theta$) can vary as well through a harmonic potential with force constant $k$ (SM). All physical quantities referring to this model are expressed in Lennard-Jones units of mass $m$, energy $\epsilon$ and length $\sigma$.

%Basic building blocks are connected pairs of beads ({\it i.e.} a dumbbell) orthogonally oriented with respect to the membrane plane in which they diffuse freely (xy-directions). The bottom beads are constrained to lay on the plane, and interact only through a volume exclusion short-range repulsion, while the top beads interact via a Lennard-Jones potential ({\it i.e.} it \textcolor{black}{also} includes an attractive tail). The z position of the top bead can assume two values (the two minima of a quartic potential); \textcolor{black}{per semplioficare dopo il linguaggio della descrizione useremo la magnetic terminilogy} these two top-bead heights can be mapped to the spin values -1 and +1. \textcolor{black}{frase su cosa succede in questo modello: at fixed dumbbell  position, witouth the tilting potential, this model is shown to a transition similar to ferromagnetic/paramagnetic transition varying the temperature.} The dumbbell vertical orientation (described by the tilting angle $\theta$) can vary as well through a harmonic potential with force constant $k$ (SM). 

%Saturated and unsaturated lipids are modeled by dumbbells with similar\textcolor{black}{altro aggettivo? equivalent? similar but not identical} characteristics (Fig.~\ref{fig1}b). 
Considering the fact that unsaturated lipids form thinner membranes and occupy on average a larger area per lipid than saturated lipids, we modelled the former by ``unsaturated'' dumbbells ($d_u$) which occupy preferentially the  s=-1 spin state and have a small force constant $k_u=70$, and the latter by ``saturated'' dumbbells ($d_s$) which occupy preferentially  the s=+1 state and have a larger value of $k_s=10^4$ (Fig.~\ref{fig1}b). Each dumbbell type has thus a typical height, although both the s=-1 and s=+1 states are accessible and occasionally populated by all dumbbells.  At any fixed temperature, constant pressure molecular dynamics simulations of single-type dumbbells show that due to the different values of $k$, $d_u$ occupy on average a larger area $a$ compared to $d_s$ (Fig~\ref{fig1}c). Importantly, both $d_u$ and $d_s$ show a transition between the $L_o$ and $L_d$ phase at a critical temperature $T^*$ (Fig~\ref{fig1}c, top panel), where  these phases are characterized by  bottom beads being hexatically ordered or disordered, respectively  (Fig~\ref{fig1}c, bottom panels). $d_u$ present a transition at $T_u^*\sim 0.41$ and $d_s$ at $T_s^*\sim 0.436$.

When $d_s$ and $d_u$ are simulated together in a 50-50 mixture, we have the additional tendency of dumbbells of same height to interact more strongly, and thus to segregate. This effect depends on the dumbbells height fluctuation, and thus is expected to be temperature-dependent. In general, within the $d_s$-$d_u$ mixture, we  have a non trivial combination of the demixing tendency and the $L_o$/$L_d$ phase transition occurring for each dumbbell type (Fig~\ref{fig1}d). Our choice of parameters led to a behavior which is consistent with the one of ternary mixtures. We find in fact that 
for $T=0.41\sim T_u^*<T_s^*$, the two dumbbell types demix, with $d_s$ forming a high-density, hexatically-ordered phase, identified as $L_o$ domains, while $d_u$ form a disordered ($L_d$) phase. 
For $T=0.436\sim T_s^*$, we still observe a nearly phase-separated system where mixing starts to occur but clusters still fluctuate in size, and both phases are disordered ($L_d$). \textcolor{black}{These fluctuations are large as expected for binary mixture near the demixing transition~\cite{stanley1999scaling}.}
For $T=0.49> T_s^*$ we have a completely mixed and disordered phase ($L_d$).

We now introduce a simplified representation of ion channels, to model the  observation that each distinct configuration (open or closed) interacts preferentially with a specific lipid species. The goal is to understand how the coupling between the internal degrees of freedom of the channel and those of the lipids can give rise to the collective phenomena mentioned in the introduction. 
\textcolor{black}{Our goal is to introduce a simple non-trivial model representing a voltage-gated ion channel. A single dumbbell has only two accessible states, up and down, while a collection of dumbbells increases the combinatorial landscape. The presence of multiple states makes it possible to study cooperativity as an enhanced two-state character of the system.}

\begin{figure*}[h]
\centering %1.5cm 0.5cm 2.3cm 9cm
\includegraphics[width=2\columnwidth]{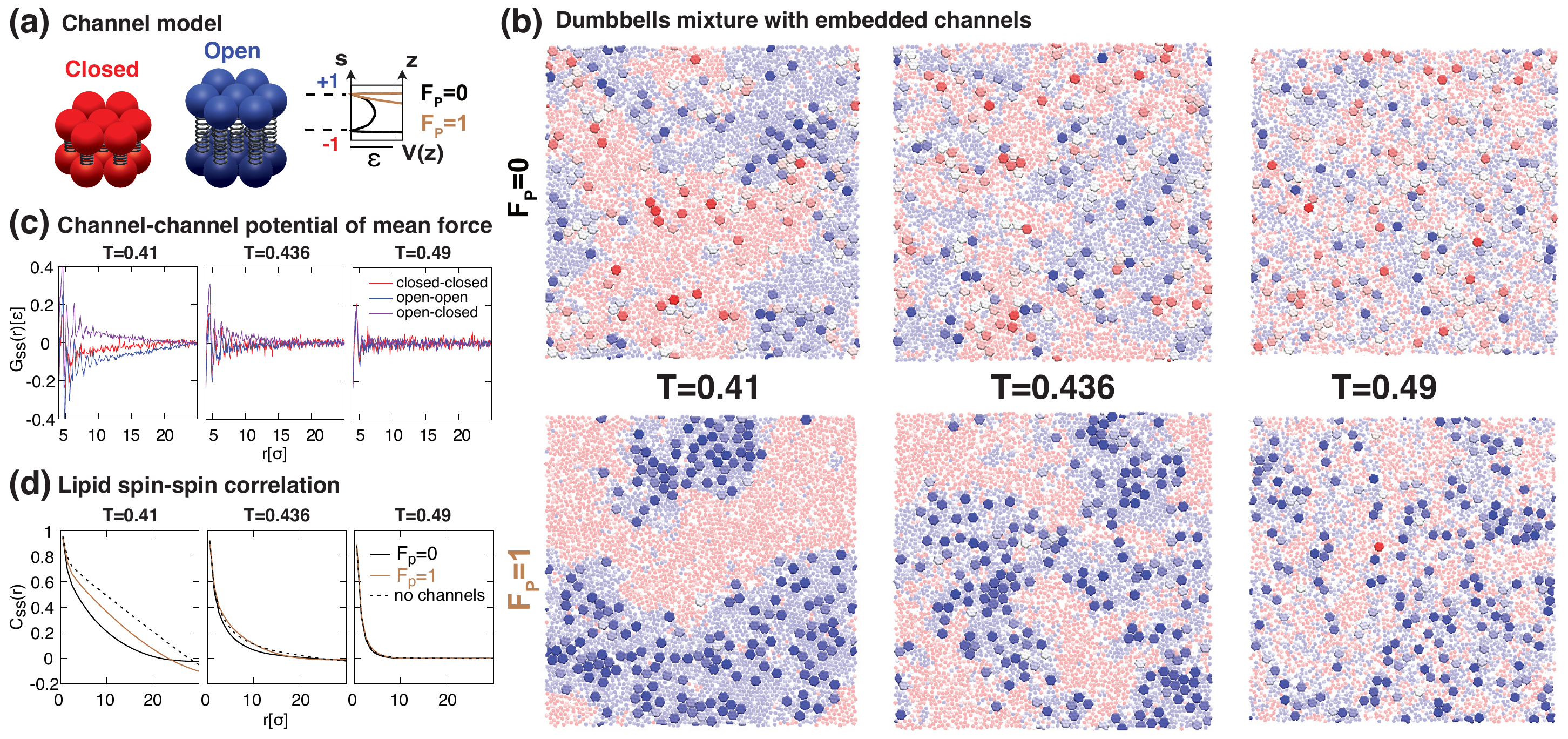}   \\
\caption{ (a) Mesoscopic model of ion channel, with seven dumbbells assembled in a hexagonal shape and co-planar neighbour beads connected via springs (SM). Top beads are subject to an external field $F_p$ in the z direction, which models the membrane potential.  Average top bead height fluctuates around two equally probable states when $F_p=0$, while one of the two states becomes favored when $F_p\neq 0$ (see potential $V(z)$ of $z$, the average channel top beads height). The central dumbbell has the role of the pore, with its spin mapped to a closed-open state. (b) Snapshots of embedded channels (darker colors) in the binary dumbbell mixture of Fig.~\ref{fig1}d, at three temperatures ($T=0.41, 0.436, 0.49$) and $F_p=0, 1$ (top and bottom row). Channel/lipid ratio is $f_p=0.05$, and dumbbells are colored according to their spin value (color scheme in Fig.~\ref{fig1}d).  (c) Potential of mean force as a function of distance between two channel pores in the open-open, closed-closed, and open-closed configuration, same $T$ as in (b) and $F_p=0$. (d) Lipids spin-spin correlation with embedded channels and compared to lipid only mixture, for the cases shown in (b).}
\label{fig2}
\end{figure*}

The channel is a collection of dumbbells arranged in a hexagonal shape, with co-planar neighboring beads connected via springs  (Fig.~\ref{fig2}a). 
\textcolor{black}{This shape does not perturb the hexatic order of dumbbells and gives a sufficiently large interaction surface with lipids.}
The $z$ coordinate of top beads is subjected to an external bias force, $F_p$, mimicking the membrane potential. Top beads move on $z$ in a two-well potential, with equally probable states without external bias ($F_p=0$) and with a smaller barrier than lipid dumbbells, to facilitate transitions between the two states. When biased, one state becomes more populated than the other. By assigning the role of ``pore domain" to the central dumbbell, we define a proxy for the two conductance states, closed and open, based on the dumbbell spin state. Accordingly, the remaining six dumbbells can be interpreted as voltage sensor domains. Importantly, due to the attractive Lennard-Jones potential, each channel state interacts preferentially with lipid species having the same height (or spin state). Channels interact with one another or with lipids via the same potential (SM).

We first look at the channel-channel clustering tendency due to lipid-mediated interactions and the resulting segregation effects. We then turn our attention to cooperativity and dynamical and out-of-equilibrium effects.

%\subsubsection{Spatial patterns of channels in lipid mixture}
{\bf Lipid-mediated interactions.}
%Fig.~\ref{fig3}b illustrates the effect of channels inside a membrane, at varying temperatures (T=0.41, 0.436, 0.49) and fixing the ratio of protein to lipids to $f_p=0.05$. We show this effect without and with an applied external potential on the proteins, obtained by applying to each top-bead of the channels a force $F_p=0, 1$ (top and bottom row, respectively, see Fig.~\ref{fig3}a for potential).
Figure~\ref{fig2}b demonstrates the effect of channels within a membrane at the three temperatures considered (T=0.41, 0.436, 0.49)  with a fixed channels-to-lipids ratio of $f_p=0.05$ (we will refer from now on to $d_s$ and $d_u$ dumbbells as lipids). The top row shows the effect without an applied external potential on the channels, $F_p=0$, while the bottom row shows the effect with $F_p=1$.

%Starting with $F_p=0$, we observe comparing  Fig.~\ref{fig3}b with Fig.~\ref{fig2}e that protein have a disordering effect on the lipid demixing at $T=0.41$. This can be explained by considering that non-biased proteins can act as surfactant between the two phases (see e.g. [straub pantelopulos surfactant exploring the impact of protein in the ternary mixture] ci sta?). Inversely, lipid channels act themselves as surfactants of proteins, which remain in a diluted regime sporadically forming small clusters. Typically, non-biased protein can diffuse in both lipid phases, and tend to assume their magnetization. Similar effect is observed at higher temperatures too. 

Comparing  the first row of Fig.~\ref{fig2}b  to Fig.~\ref{fig1}d, we see that channels promote lipid mixing at $T=0.41$ and $F_p=0$, in the sense that the $L_d$ and $L_o$ interface margins are less well-defined. This can be explained by the fact that unbiased channels can act as lineactants between the two phases~\cite{bandara2019exploring}. On the other hand, the lipid-channel attraction strength is the same as for channel-channel, and thus channels remain in a diluted state.  Across all temperatures, unbiased channels can diffuse into both lipid phases and tend to assume their height or spin state. 
%This effect is also observed at higher temperatures. \textcolor{black}{a cosa ci riferiamo? non si capisce. togliere}

% This is mainly due to channels acting as surfactants for the lipid rafts similarly to ref [straub pantelopulos surfactant exploring the impact of protein in the ternary mixture]. In a similar way, protein at this density should phase-separate when left alone, thus lipids inhibit condensation of proteins with a similar process (check). 
%The phase looks similar to the one at critical temperature without channels, and increasing temperature gives a similar behavior  (mixing of phases) as the one observed without proteins (Nota si potrebbe vedere che succede per temperatura piu piccola all esatico delle due fasi, si dobrebbe essere abbassato)

%When an external field is applied to proteins, these are biased towards one or the other magnetization, based on the direction of the force. Here we show $F_p=1$, which bias protein to have a spin-up.
%Notably, at equilibrium protein migrate in the favored phase, which here is the one with spin-up lipids. In this case, they effectively act as lipids of one phase, thus the system behaves more similarly to one of only lipids (compare again Fig.~\ref{fig3}b with Fig.~\ref{fig2}e).

When an external field is applied to channels, they are biased towards one or the other spin state based on the direction of the force (Fig.~\ref{fig2}a), assuming either an open or closed conformation. In the case of Fig.~\ref{fig2}b, we apply a force of $F_p=1$ which biases the channels to adopt the spin-up (open) conformation.
At equilibrium, channels migrate towards the phase with the same height, which in this case is the phase with spin-up lipids. As a result, channels effectively behave like lipids of the same height, causing the system to become more similar to one made up of only lipids (compare second row of Fig.~\ref{fig2}b to ~\ref{fig1}d).

%b) We now look at the effect of adding a bias potential: protein will behave and have a much similar state to proteins, which induce them to migrate in the favoured phase if present (red if V=-1 and blue if V=1). The second visual effect is that, behaving more as lipids, they contribute to the ordering of the lipid themselves. In fact, at T=0.41 the phase becomes visually more similar to the one of lipid alones.

%Of particular interest is observing if proteins display some attraction/repulsion effect simply by being immersed in the same/opposite lipid state, similarly to what predicted theoretically in Ref.~\cite{machta2012critical}.
%To do so, we compute the potential of mean force between proteins if they are parallel/antiparallel (Fig.~\ref{fig3}c). Indeed, we observe that parallel proteins display long-range attraction, which is higher near the demixing transition, and decreases when lipids start to mix, while antiparallel proteins are repulsive towards each other. This effect is easily understood by thinking of proteins in the same phase to be constrained to have the same spin, independently of their distance. Thus, the typical decay observed in the effective potential reflects the size of the underlying lipid patches.

Of particular interest is the determination of whether channels show an attraction or repulsion between each other, \textcolor{black}{due exclusively to the fact that channels are} immersed in  lipids with the same or opposite spin.
To investigate this effect, we calculated the potential of mean force between channels when they are both open or both closed and when one is open and the other one is closed (Fig.~\ref{fig2}c). \textcolor{black}{The potential of mean force is computed as $G_{ss}(r)=-k_BT\ln{P_{ss}(r)}$, with $P_{ss}(r)$ being the probability of having two channel pores with spin values $ss$ at distance $r$.} We find that channels in the same conformation display long-range \textcolor{black}{lipid-mediated} attraction, which is stronger near the demixing transition and decreases as the lipids begin to mix. Moreover, channels in the open conformation will feel a stronger attraction than channels in the closed conformation. In contrast,  channels in different conformations  \textcolor{black}{show long-range lipid-mediated repulsion}. These effects can be understood by considering that channels in the same phase are constrained to have the same conformation, regardless of their distance.  Open channels are embedded in a denser environment and, as a result, lipid-mediated interactions are stronger. The range of the potential of mean force reflects the typical size of the underlying lipid patches. \textcolor{black}{Note that the existence of these lipid-mediated long-range interactions occurring under critical conditions (known as Casimir forces) was already predicted theoretically~\cite{reynwar2008membrane,machta2012critical} and shown in atomistic simulations~\cite{reynwar2008membrane,katira2016pre}.}
%In funzione della temperatura the presence of the underlying bilayer gives rise to an effective attraction between proteins immersed, which is simply due to proteins being immersed in the same raft, and thus constrained to be in the same conformation. The decay in the effective potentail follows the one observed for the lipid bilayer model at the same temperature (dalla lunghezza tiica delle patches) This is the combined effect of segregation, which promotes aggregation, and the effect due to casimir forces che attraggono gli spin uguali se sono nello stesso raft.  (Nota check senza attrazione e si potrebbe fare vdere a diversi bias che succede e commentare. Faccio un check a T=0.41 variando la densita, e uno con N=204 al variare della temperatura. Questo check andrebbe fatto anche per vedere che cambia con l'activation curve- Ho controllato che al variare della temperatura c'è attivazione che decresce. Ho controllato che la forza di casimir è praticamente uguale senza attrazione. E' dovuta alla presenza dei lipidi che fanno da intermediari.

Conversely, the effect of channels on the lipid patch can be understood by considering the lipid-lipid spin-correlation function (Fig.~\ref{fig2}d)\textcolor{black}{, defined as $C_{ss}(r)=\langle s_i s_j\rangle$, with $s_i$ the spin assigned to each lipid, and $r$ the distance between two lipids}. At $F_p=0$ and temperatures near the demixing, the function decays much faster than the corresponding decay calculated for systems without channels (dashed line), while at $F_p=1$ the correlation becomes longer. \textcolor{black}{The longer-range correlation for $F_p=1$ is a consequence of the fact that channels are all in the same state and thus attract same-height lipids around them, locally increasing the lipid spin-spin correlation. Conversely, at $F_p=0$ channels are free to fluctuate between the two states and adapt to the local environment. This lineactant-like behavior causes a  roughening of the domain boundaries and thus a decrease in lipid spin-spin correlation. }

%An effect which is instead observed in the lipid themselves, is that the correlation function for V=0 is much smaller than the corresponding one observed for lipids alone. As soon as protein behaves as single-state molecules, and incorporetes themselves in one phase, they stabilize it and improve the lipid correlation back to regimes more compatible with the non-protein behavior.

\begin{figure*}[h]
\centering %1.5cm 0.5cm 2.3cm 9cm
\includegraphics[width=1.8\columnwidth]{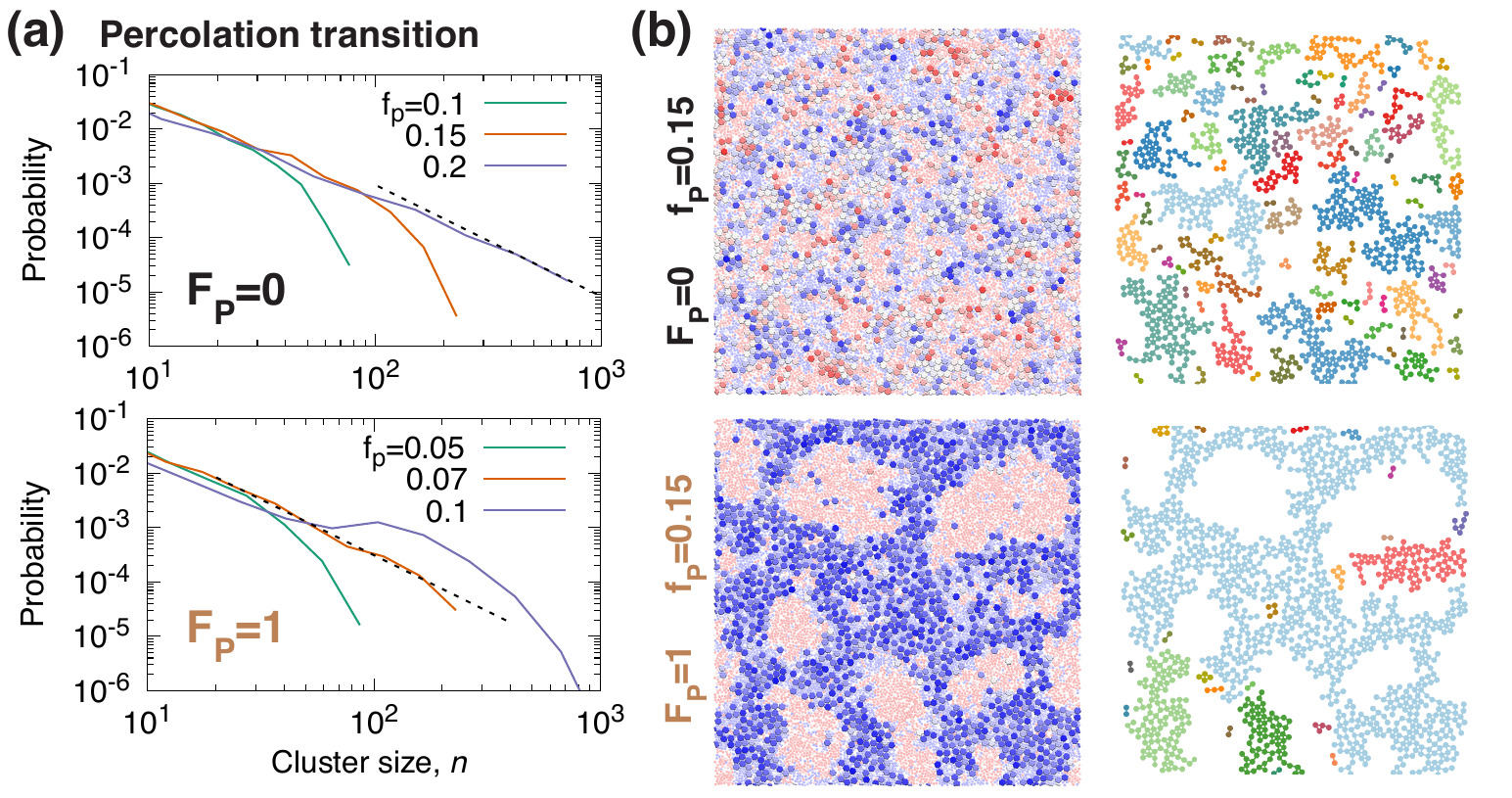}   \\
\caption{ a) Cluster size distributions at $T=0.436$, $F_p=0, 1$, and different channel concentrations $f_p$. Upon increasing $f_p$, the distribution changes from a power law with an exponential cut-off, to a pure power law at the percolation transition, to a single peak (not shown). The critical density decreases from $f_p\sim 0.2$ at $F_p=0$ to $\sim 0.07$ at $F_p=1$. (b) First column: Spin-colored snapshots of system with $f_p=0.15$, $T=0.436$, at $F_p=0, 1$ (top and bottom row). Second column: same system configurations, showing only the pore dumbbells colored according to the cluster's label. Despite the fact that $f_p$ is the same, at $F_p=1$ channels are more interconnected than at $F_p=0$, as they are pushed in the spin-up phase.}
\label{fig3}
\end{figure*}

{\bf Percolation transition.}
We then turned our attention to the effects of channel density on the system, Fig.~\ref{fig3}, fixing $T=0.436$. At $F_p=0$ and for small values of $f_p$, we observe that the size distribution of channel clusters has the typical shape of a power law with  exponential cut-off correction, $P(n)\sim n^{-\tau} e^{-n/n^*}$, with $n^*$ a typical cluster size (Fig.~\ref{fig3}a, first row). By increasing the channel density, the distribution first tends to a pure power-law for $f_p\sim 0.2$, $P(n)\sim n^{-\tau}$, and then it develops a peak at large $n$ (not shown). This behavior is in general associated with a percolation transition~\cite{stauffer2018introduction}, which in two dimensions is characterized by the critical Fisher exponent $\tau=187/91$. Notably, the same exponent is observed in the experiments shown in Fig.~S1 from Ref.~\cite{sato2019stochastic}. %What occurs is that protein start to gradually cluster at increasing density, until all protein come into contact across the boundary conditions, forming a single large cluster.

An interesting role in this percolation transition is played by the external field $F_p$ (Fig.~\ref{fig3}a, second row). When applied, the field shifts the transition towards lower densities, $f_p\sim 0.07$. The reason is that, when activated, channels are encouraged to migrate to the most favorable lipid phase. In this way, the surface area occupied by channels decreases (by a factor of 2) and, accordingly, the effective density increases (Fig.~\ref{fig3}b). Overall, the percolation transition is observed for a smaller channel-lipid ratio.

{\bf Channel-Channel coupling and cooperativity }
%Our next We now turn to characterize if, due to presence of lipid-mediated interactions, there are any effects in the channel response. The latter is a typical measurement done in patch-clamp experiments with the charge output of channels Q measured as a function of the applied voltage V. Here, as a proxy to reproduce such curve we computed the probability of particles being spin-up, $P_{up}$, as a function of the bias force, $F_p$, Fig.~\ref{fig5}a. 
We now investigate whether the presence of lipid-mediated interactions has any effect on the ion channel response to the trans-membrane potential. Experimentally, this may be investigated in a voltage-clamp setting by recording the \textcolor{black}{steady-state ion current, or} pore conductance, $G$, as a function of the applied voltage $V$. \textcolor{black}{Voltage-gated ion channels contain specialized domains (voltage sensors) that increase the voltage sensitivity of the conduction pore through allosteric coupling; this increased responsiveness to the transmembrane potential is manifested as a steepening of the “G-V” curve. We expect lipid-mediated interactions between voltage sensors in the same channel and between neighboring channels to enhance voltage sensitivity of conductance in a predictable temperature-dependent manner.} 

\begin{figure*}[h]
\centering %1.5cm 0.5cm 2.3cm 9cm
\includegraphics[width=1.8\columnwidth]{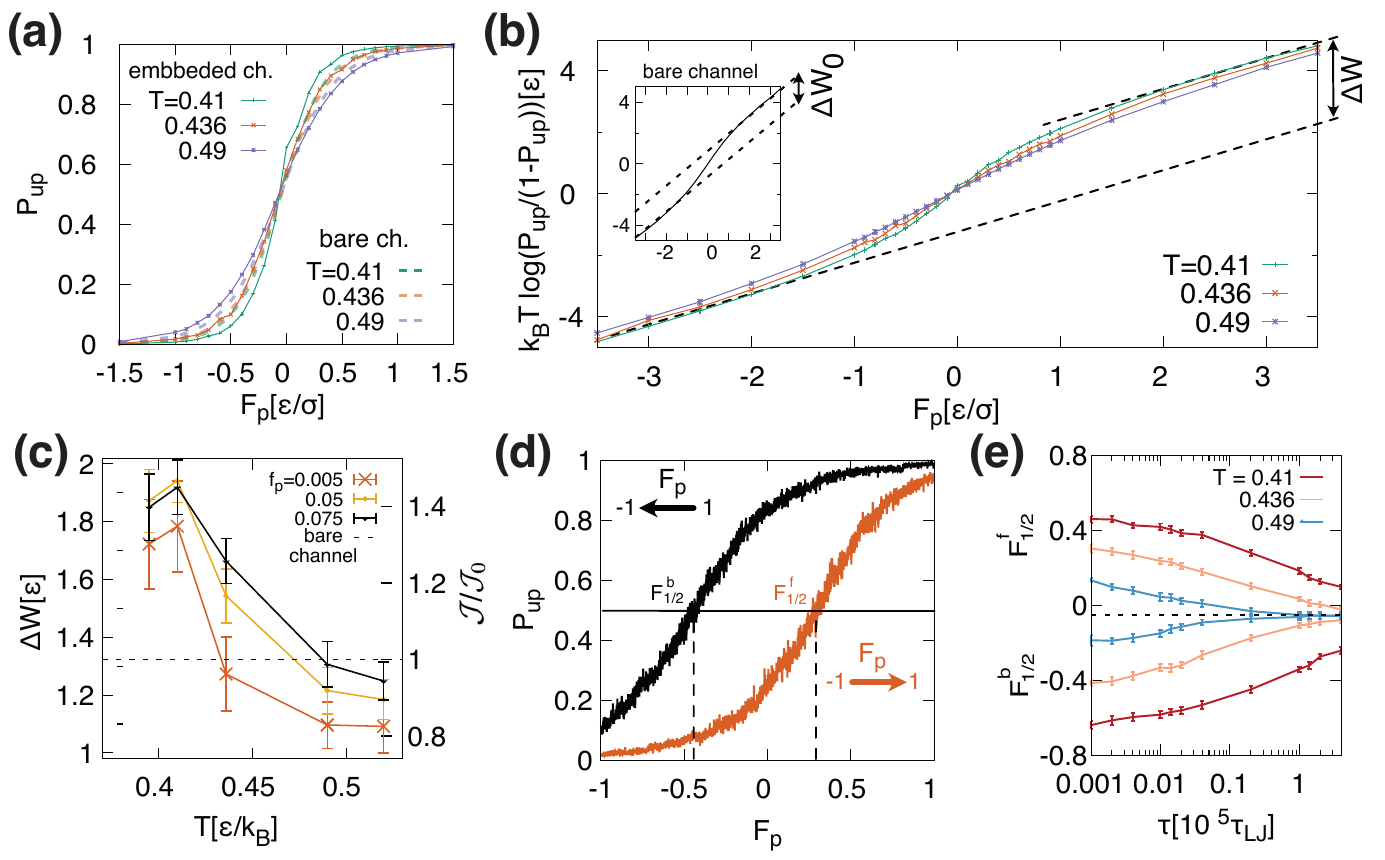}   \\
\caption{(a) Probability that the pore dumbbell of a channel\textcolor{black}{, embedded in the lipid bilayer,} is in the up-state ($P_{up}$) as a function of the external field $F_p$, for different temperatures, at $f_p=0.05$ \textcolor{black}{ (continuous lines). For comparison, the same curves for a bare channels are plotted (dashed lines) at the same temperatures. These are  shifted on the $F_p$ axis so that the values at $P_{up}=0.5$ coincide with those of the embedded channels}. (b) Hill plot of the same data shown in (a). The  pore linkage energy, $\Delta W$, defined as the  vertical separation between the Hill's plot asymptotes, decreases upon increasing of the temperature. In the inset, the Hill plot for a bare channel is shown. (c) $\Delta W$ as a function of temperature for three different densities $f_p$. The dashed line shows the temperature independent reference value $\Delta W_0$ of the bare channel. (d) Activation curves  at $T=0.41$, $f_p=0.05$ obtained by ramping-up from $F_p=-1$ to 1 (forward, red), and viceversa (backward, black), in a timing interval $\tau=2\cdot 10^4\tau_{LJ}$. Note the marked hysteresis. Field strengths required to activate half  of the channels ($P_{up}=0.5$) are indicated as $F_{1/2}^f$ and $F_{1/2}^b$  for the forward and backward processes, respectively. (e)  $F_{1/2}^f$ and $F_{1/2}^b$ as a function of the ramp duration $\tau$. \textcolor{black}{Averages are performed over 5 independent runs.} } 
\label{fig4}
\end{figure*}

To reproduce \textcolor{black}{the “G-V”} curve, we computed the \textcolor{black}{equilibrium} probability, $P_{up}$, that the central "pore" particle of each 7-dumbbell channel is in a spin-up state (open conformation). \textcolor{black}{The conductance is related to $P_{up}$ through $G=NgP_{up}$, where $N$ is the number of channels and $g$ is the single pore conductance (assumed constant)}. We simulated a normalized ``G-V" curve by calculating $P_{up}$ as a function of the bias force, $F_p$, as shown in Fig.~\ref{fig4}a\textcolor{black}{, where we compare these curves in the case of an isolated (bare) channel, and of a channel embedded in the lipid bilayer, for three different temperatures. As anticipated, we notice a temperature dependence, which is stronger in the case of embedded channels.} 

However, one must distinguish between the effect due to a trivial temperature dependence of Boltzmann weights and the temperature dependence that arises from subtle modulations of  intra- and inter-channel interactions. To discriminate between these two effects, we considered the ``conductance" Hill plot, which eliminates the trivial $T$-dependence by reporting on the free energy of pore activation $W = k_BT\log(P_{up}/(1-P_{up}))$ as a function of the external field~\cite{sigg2013linkage}. 
\textcolor{black}{ In particular, we are interested in the vertical separation $\Delta W$ between the two linear asymptotic trends observed at extreme field strengths. This amounts to removing from the free energy the linear field-coupling term, thereby isolating the effective interaction energy between the pore and the voltage-sensitive particles~\cite{sigg2013linkage}. The entire procedure eliminates the weak inverse-temperature dependence of the “G-V” slope intrinsic to the Boltzmann distribution, which exists even for the bare channel (Fig.~\ref{fig4}a), and reveals a more interesting lipid-mediated temperature sensitivity.}

%We investigated the dependence of the "G-V" curve on temperature. In Fig.~\ref{fig4}a, the activation steepens with decreasing temperature. However, one must distinguish between the effect due to a trivial temperature dependence of Boltzmann weights and the temperature dependence that arises from subtle modulations of  intra- and inter-channel interactions. To discriminate between these two effects, we considered the ``conductance" Hill plot, which eliminates the trivial $T$-dependence by reporting on the free energy of pore activation $W = T\log(P_{up}/(1-P_{up}))$ as a function of the external field~\cite{sigg2013linkage}. We define the coupling energy $\Delta W$ as the vertical separation in free energy between linear asymptotic regions of $W$ at large positive or negative applied fields. 

An exact expression of $W$ is found for the isolated (bare) 7-dumbbell channel, which exists as a system of connected spins with ferromagnetic coupling constant $\mathcal{J}_0$ between the pore and each voltage sensor (SM). In this system, the coupling energy of the bare channel is found to be proportional to $\mathcal{J}_0$, $\Delta W_0=24\mathcal{J}_0$. Embedding the channels in a lipid environment affects the coupling energy $\Delta W$, see Fig.~\ref{fig4}b. This modulation of lipid-channel and channel-channel interactions can be captured by an effective,  temperature-dependent coupling constant, $\mathcal{J}\equiv \Delta W/24$, which takes into account the effect of the environment  and provides a quantitative measure of the system's cooperativity. Increasing $\mathcal{J}$, and thus $\Delta W$, steepens the closed-to-open transition emphasizing the two-state character of the channel.

Fig.~\ref{fig4}c shows the fitted $\Delta W$ for different temperatures and channel densities, compared to the temperature-independent value $\Delta W_0$ for the bare channel (SM for the fitting procedure). Two major observations can be made from the figure. 

First, there is a crossover between $\mathcal{J}$ and $\mathcal{J}_0$ around the critical temperature ($T\sim 0.436$) with $\mathcal{J}>\mathcal{J}_0$ or $\mathcal{J}<\mathcal{J}_0$ depending on whether the channel is embedded in a demixed or mixed lipid membrane. This dependence on temperature highlights the crucial role of membrane phase behavior. In the lower temperature demixed state, channels can migrate toward the lipid patch that stabilizes the instantaneous conformational state (i.e. closed or open). This is not possible in the higher temperature mixed state, in which the fluctuating local lipid composition gives rise to a more heterogeneous ensemble that decreases the overall two-state character of the system. 

Second, higher channel densities ($f_p$) result in larger $\mathcal{J}$ at a given temperature, suggesting that channels are coupled to one another and tend to activate in a cooperative fashion; the biggest gap occurs around the demixing temperature, $T\sim 0.436$. We infer that channel-channel interactions (both direct and lipid-mediated) increase cooperative gating beyond what is expected from independently-gated channels. \textcolor{black}{Notably, similar channel-channel gating cooperativity was found in mechanosensitive channels~\cite{paraschiv2020dynamic} where their attraction promotes channels closure.}

%The dependence of $J$ on the temperature, namely point i), highlights the importance of the lipid environment: lipids can in fact have an important effect while surrounding proteins, which can be positive or negative depending on their type of phase. A demixed phase allows lipids to easily maintain their open/closed state depending on which phase they are, while a disordered environment has the opposite effect of making the protein state more uncertain. 

%The dependence of $f_p$ of point ii) is important as well: protein-protein interactions, if present, allow them to have a more robust and cooperative response to an external field, as they act together, while surrounding lipids do not have a strong response to the field. Moreover, we know that the membrane lives near the demixing transition xxxrefxxx, which might be linked to having a stronger coupling xxx che altro?xxx.

{\bf Out-of-equilibrium effects and hysteresis.}
The demonstration that channel-lipid interactions enhance cooperative gating raises an interesting question: since the response of any given channel depends crucially on the activation state of neighbor channels (and on the thermodynamic phase of the surrounding membrane), is the closed-to-open transition quasi-static or dominated by non-equilibrium effects? 
%What we have shown above suggests that one of the possible causes of cooperativity in ion channel activation is the energetic interaction that couple channels  with lipids and thus with one another. A question that emerges naturally is then whether or not a dynamical coupling is present as well. Since the response of any given channel depends crucially on the activation state of the remaining ones (and on the thermodynamic phase of the surrounding membrane), is the closed-to-open transition adiabatic or dominated by out-of-equilibrium effects? 

To answer this question we characterized the response of channels to a time-dependent external field with a strength that increases linearly with time. Fig.~\ref{fig4}d illustrates $P_{up}$ as a function of the instantaneous value of the applied $F_p$. The system starts at equilibrium with $F_p=-1$ and the field is gradually increased to $F_p=1$ for a total ramp duration $\tau$ (red curve, ON ramp protocol); the black curve shows the mirror-image OFF ramp protocol. A microscopically reversible process should give rise to overlapping curves. Instead, there is definite hysteresis even for relatively long ramp times $\tau$. Close inspection of a sample trajectory highlights the reason underlying this behavior. Inverting the polarity of the applied field results in two combined effects: it causes channels to migrate from one phase to the other, and it promotes a phase transformation in the region surrounding channels. If $\tau$ is small compared to both the typical diffusion time $L^2/D$ (where $L$ is the typical patch size and $D$ is the channel diffusion constant) and the phase-separation relaxation time, then the process is out of equilibrium and hysteresis appears. 
To quantitatively characterize this behavior, we report the field value resulting in the activation of half of the channels, namely $F_{1/2}$, as a function of $\tau$ for ON and OFF ramps (Fig.~\ref{fig4}e). We note that the ON and OFF protocols tend asymptotically to the same equilibrium value and that the rate of convergence is markedly temperature-dependent, with hysteresis being stronger for lower temperatures. 

\textcolor{black}{Regarding finite size effects, one expects that the system size imposes a cutoff on the maximum patch size encountered, and in turn the latter changes the characteristic time needed for channels to cross different patches. We thus expect that dynamical properties are affected by finite size effects more than equilibrium ones. Hysteresis, which 
depends on the time required for channels to migrate from one patch to the other is indeed enhanced by considering larger systems (see SM).}

\subsection{Bridging membrane physics and physiology: a lattice model}

\textcolor{black}{We shift gears from the molecular dynamics model that is chiefly concerned with the phase structure of the membrane to a system with more physiological relevance.} Physiologists have traditionally modeled ion channel gating kinetics using continuous-time discrete-state Markov models~\cite{Sigworth1994Review, Bezanilla2018Review}. Gating schemes with a finite number of states adequately fit macroscopic and single-channel data in response to a variety of external stimuli~\cite{MCMANUS1988Markov}, but may be inadequate in describing types of anomalous gating explored in this paper---clustering-induced cooperativity, hysteresis, modal gating, and long-term memory. The crux of our thesis is that a complex membrane absorbs and distributes information like a field, altering the kinetics of bare channels. We thus set out to include additional degrees of freedom to the traditionally used discrete-state models to account for lipid-channel coupling. We demonstrate that this paradigm can provide a unified explanation for anomalous gating.

\begin{figure*}[h]
\centering %1.5cm 0.5cm 2.3cm 9cm
\includegraphics[width=2\columnwidth]{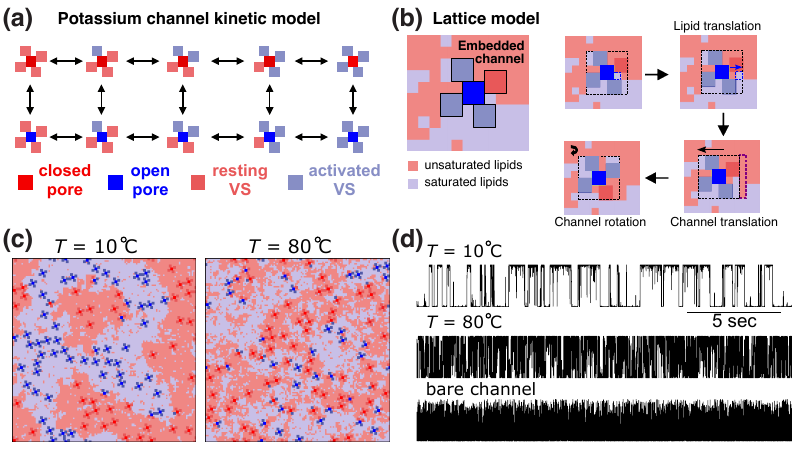}   \\
\caption{(a) 10-state activation scheme for a potassium channel. \textcolor{black}{The central pore, in dark red/blue for closed/open state, is surrounded by four allosterically linked voltage sensors (VS) in light red/blue for resting/activated states. Rate constants between different states are described in the SM.} 
%The states are labeled with powers of the equilibrium constants \textit{P} and \textit{J}. The statistical weights for each state are easily obtained from a binomial expansion of the partition function \textit{Z}. Intrinsic transition rates $\nu_P$ and $\nu_J$ are 2 kHz.  
\textcolor{black}{(b) Lattice model. The channel is embedded as 
 (2×2) pore in contact with four symmetrically arranged (2×2) voltage sensors. The channel is surrounded by embedded lipids (ligther red/blue for unsaturated/saturated lipids) occupying each a single cell. Implemented translational and rotational degrees of freedom are shown on the right.} (c)  Snapshots of the lattice model at $T=10^{\circ}, 80^{\circ}$C, and approximate half-activation voltage $V=40$mV below and above the demixing transition. (d)  Conductance time course of a single embedded channel at  $T=10^{\circ}, 80^{\circ}$C, and bare channel kinetics at $T=30^{\circ}$C. }
\label{fig5}
\end{figure*}

%To test our hypothesis in a physiologically relevant context, 
\textcolor{black}{
We consider an established kinetic model for potassium channel gating and extend it to include the effect of channel-lipid interactions. A channel system suitable for our purposes is the large-conductance calcium-activated potassium (BK) ion channel, which, under conditions of zero- or saturating calcium, is regulated by four identical voltage-sensing domains in a manner closely described by the Monod–Wyman–Changeux (MWC) model of allosterism\cite{Cox1997BK,Horrigan1999_BKzeroCa,shelley2010coupling}. The kinetic model corresponding to an MWC scheme contains ten states (Fig~\ref{fig5}a). Each forward step corresponds to the activation of a single domain, either the pore or one of the voltage sensors. This discrete-state model results in a more realistic description of gating kinetics and a more faithful structural representation than could be achieved with the hexagonal model studied with molecular dynamics.
}

\textcolor{black}{
In order to study the effect of a dynamic lipid environment, we constructed a square-lattice system that best adapts to the four-fold symmetry of the BK channel and possesses Ising-like interactions between neighboring binary cells, see Fig~\ref{fig5}b for a schematic representation. The lattice contains channels and lipids, of which there is a 50:50 mix of two types, “saturated” and “unsaturated”. The complete lattice is large enough ($128^2$ cells) to contain 100 channels.  An individual channel is composed of a central pore in contact with four symmetrically arranged voltage sensors, occupying a total of 20 cells. The activation of each voltage sensor corresponds to the transport across the membrane of an elementary charge q. Macroscopically, this gives rise to a current whose time integral Q is a typical measurement done in patch-clamp experiments. The response of Q to the applied voltage V is encoded by the "Q-V" curve.
Each lipid, which occupy a single cell, is permanently fixed in the ``up'' or ``down'' state depending on whether it represents a saturated or unsaturated species. Channels can translate and rotate, while lipids only exchange their state with neighbouring lipids (right side of Fig~\ref{fig5}b), so that the  saturated/unsaturated ratio is preserved. Lipids are mobile and preferentially associate with their own type via an energy penalty incurred for oppositely aligned pairs. 
Importantly, lipids interact with the pore and the voltage sensor in a state-dependent fashion with the same energy penalty, where unsaturated/saturated lipids preferentially bind the resting/activated states. 
}

\textcolor{black}{
The model parameters were chosen to be consistent with broad stroke features of ion channel dynamics, namely i) the steepness and relative positions of equilibrium gating charge (Q-V) and conductance (G-V) curves~\cite{Horrigan2002BKgating,sigg2013linkage}, ii) out-of-equilibrium decay rates of these same quantities~\cite{shelley2010coupling}, and iii) translational diffusion coefficients of lipids and channels. The dynamics of the system were simulated using the kinetic Monte Carlo method of Gillespie~\cite{Gillespie1977} based on the chemical master equation whose rate constants from two states are assumed to have an Arrhenius form with a different rate coefficient for each specific Monte Carlo move. A detailed description of parameter choices and methods for computation can be found in the SM.
}

Fig.~\ref{fig5}c shows a snapshot of the lattice system near the mid-point of channel activation ($V = -40$ mV) at sub-critical (10 $^{\circ}$C) and supra-critical (80$^{\circ}$C) temperatures. At 10 $^{\circ}$C the system is phase-separated (demixed). Similar to the mesoscopic model, the activation state of a channel generally aligns with the lipid phase in which it resides. At 80 $^{\circ}$C, phases are mixed. Fig.~\ref{fig5}d shows a long (20 sec) single channel tracing characterized by periods of substantially different open probabilities linked to slow transitions between different lipid phases, increasing in frequency with rising temperature. Because these phase-like transitions are slow compared to the millisecond Markovian kinetics of the isolated channel (bottom trace, at 30 $^{\circ}$C), they can be considered modal events ~\cite{PATLAK1979Modal,Magleby1983Modal}.

\begin{figure*}[h]
\centering %1.5cm 0.5cm 2.3cm 9cm
\includegraphics[width=2\columnwidth]{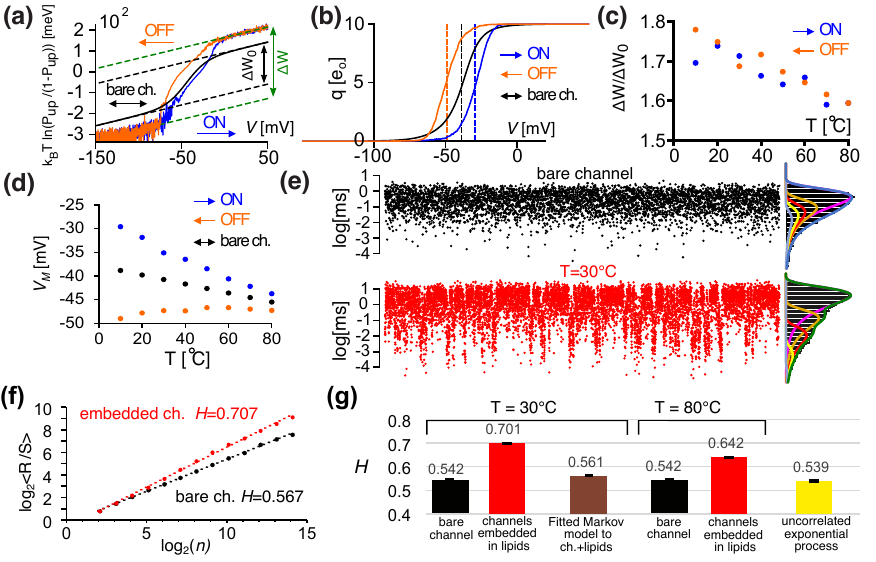}   \\
\caption{(a) Hill plot for 2-second rising ON (blue) and  falling OFF (orange) voltage ramp ($\pm 0.1$ mV/ms) conductance of the embedded channel compared to the quasi-static behavior of the bare channel (black line), at $T=10^{\circ}$C. \textcolor{black}{Arrows indicate the ramp direction.} The pore linkage energy $\Delta W$ is the vertical separation between skew asymptotes (dashed lines). \textcolor{black}{Note the smaller separation $\Delta W_0$ for the bare channel}.  (b) Q-V curves obtained with the same ramp protocol as in (a).  The median voltages of activation $V_M$ \textcolor{black}{(values at which half of the channels are activated)} are indicated by the dashed lines. (c) Ratio of $\Delta W$ to the constant value $\Delta W_0$ \textcolor{black} { = 200 meV} of the bare channel as a function of temperature \textcolor{black}{for the ON and OFF voltage ramps}. (d) \textcolor{black}{$V_M$ values for the ON and OFF ramps as a function of temperature, compared to the bare channel case. Q-V curve hysteresis decreases with temperature as demonstrated by the convergence of the ON/OFF $V_M$ values.} (e) Sequential open dwell times for bare (black) and embedded (red) channels at $T=30^{\circ}$C, $V = 40$mV. The semi-log dwell-time distributions fitted to the eigencomponents (solid lines) of the 10-state model are shown to the right of the sequences. (f) Single-channel Hurst plots comparing slopes ($H$) of bare (black) and embedded (red) channels at $T=30^{\circ}$C. (g) Averaged ($n$ = 100) Hurst exponents $H$ for bare (black bars) and embedded (orange bars) channels at $T=30^{\circ}, 80^{\circ}$C. The two control cases with exponents near $0.5$ are: (1) a bare channel whose Markov gating scheme was fitted to the dwell time distribution of the embedded channel (brown bar); (2) an uncorrelated sequence of exponentially-distributed dwell times (yellow bar).}
\label{fig6}
\end{figure*}

{\bf Cooperativity and hysteresis}
%We now demonstrate that the lattice channel modeling a generic potassium channel exhibits similar behavior to the 7-dumbbell mesoscopic model. 
To test for hysteresis, 
we implemented a double ramp protocol with rising (ON) and falling (OFF) phases acquired at a ramp speed of $\pm 0.1$ mV/ms. With this slow-ramp protocol, entire (quasi-static) ON and OFF activation curves were acquired in a single 4-second simulation. We measured the conductance Hill plots and Q-V curves of averaged activation curves from 100 embedded channels (channel density of $f_p=7.8\cdot 10^{-3}$), Fig.~\ref{fig6}a-b. The vertical separation of skew asymptotes in the Hill plot was used to measure the pore-voltage sensor coupling energy $\Delta W$ at $T=10^{\circ}$C, for isolated (bare) and embedded channels.

The bare channel yielded overlapping ON and OFF activation curves while Q-V hysteresis was seen with embedded channels, accompanied by temperature-sensitive cooperativity (Fig.~\ref{fig6}c-d). These behaviors are consistent with the idea developed in the previous section that membrane demixing promotes channel-channel cooperativity and hysteresis arises from the relatively long time it takes a channel to migrate between phases. Increasing the temperature, we again observed lessening hysteresis from the mixing of lipid phases.
%The lipid diffusion rate does not appear to be rate-limiting; rather, the major impediment is the reluctance of channels to cross phase boundaries unless there is sufficient voltage bias to make it energetically favorable (see supplementary information for ramp cine files that demonstrate this point). Increasing the temperature decreases hysteresis by mixing lipid phases. 

%To demonstrate cooperativity between lattice-embedded channels, we implemented a 4 sec double ramp protocol with same-duration rising (ON) and falling (OFF) phases for an effective ramp speed of $\pm 0.1$ mV/ms. Hill plots of the summed conductance from 100 channels were constructed for the ON and OFF ramps (Fig. 6b), and the coupling energy $\Delta W$, defined as the vertical separation of skew asymptotes, was determined in each case and compared to the isolated Markov model, the latter demonstrating no significant differences between ON and OFF phases. The added coupling $\Delta \Delta W$ from lattice insertion decreased with increasing temperature (Fig. 6g), consistent with the idea that phase separation promotes channel-channel cooperativity.

% The ramp Q-V of the isolated Markov model was essentially identical to the true Q-V calculated from the ten-state partition function (SI), but there was marked hysteresis between ON and OFF ramps for the embedded channel (Fig.~\ref{fig6}f) due to the between-phases transition time lagging behind the ramp speed. The hysteresis decreased with increasing temperature (Fig.~\ref{fig6}h) and the ensuing decrease in average domain length.

{\bf Long-term memory and Hurst analysis.}
We questioned whether lipid-embedded channels can exhibit long-term memory of the type observed in single channel records from BK channels~\cite{VARANDA2000Hurst,Siwy2001HurstBK,Bandeira2008HurstBK,wawrzkiewicz2017temperature,Wawrzkiewicz2018HurstBK}. This phenomenon differs from short-term memory in Markov models, in which dwell time distributions demonstrate a finite number of exponential decays ~\cite{Shelley2010BKHighCa}. Long-term memory follows a power law characterized by self-similarity in a process $B(t)$ satisfying $B(at)\sim |a|^HB(t)$, where $H$ is the Hurst exponent~\cite{beran1994statistics}. If $H>1/2$ ($H<1/2$), the process has long-term positive (negative) autocorrelation, while $H=1/2$ for uncorrelated processes.

We applied Hurst analysis to consecutive open dwell times from 100 embedded channels (Fig.~\ref{fig6}e). Simulations were run under equilibrium conditions at the half-activation potential $V = -40$ mV and at temperatures $T = 30^{\circ}$C and $80^{\circ}$C, until every channel experienced at least $2^{14}$ opening events.
Bare channels governed by the ten-state Markov model demonstrated longitudinal homogeneity in their dwell-time sequence, whereas dwell-time sequences of embedded channels were more erratic. It was possible to fit the broadened dwell-time distribution of the embedded channel to the original ten-state Markov model by adjusting kinetic parameters. The fitted Markov model of the embedded channel was used as a control in the Hurst analysis.

We performed the standard determination of the Hurst exponent from the slope of $\log_2 {\rm R}(n)/{\rm S}(n)$ versus $\log_2(n)$, averaged over subsets of data of size $n$, where ${\rm R}(n)$ is the range of the summed deviation from the mean dwell time and ${\rm S}(n)$ is the standard deviation~\cite{beran1994statistics}. The slope was estimated through linear regression for $n$ ranging from 2 to 14.
This method overestimates $H$ due to sampling bias in small bins~\cite{Bassingthwaighte1994Hurst} but is consistent with previous analysis of experimental data~\cite{silva2021memory}. In practice, determining the absence of long-term correlations was achieved by comparing to internal controls.

Typical Hurst plots and fitted exponents are shown in Fig.~\ref{fig6}f-g. The average Hurst exponent for the bare channel (Markov case) is equal to $0.542 \pm 0.018$ (mean $\pm$ std. dev.), a result that is independent of temperature. Although $H$ is slightly greater than 0.5, it is statistically no different from an uncorrelated series of exponential deviates ($H = 0.539 \pm 0.019$, \textit{p} = 0.24, Student's t-test), demonstrating the absence of long-term memory.

On the other hand, in embedded channels we find $ H = 0.701 \pm 0.018$, consistent with outside experimental values of $H$ in the BK channel that vary from 0.61 to 0.71 ~\cite{VARANDA2000Hurst,Siwy2001HurstBK,Bandeira2008HurstBK,Wawrzkiewicz2018HurstBK}. There is lipid-induced broadening of dwell-time distributions that accompanies increased $H$ values, but this does not explain long-term memory, as the fitted Markov model to the embedded channel distribution is near the uncorrelated value ($H = 0.561 \pm 0.016$).
Increasing the temperature to $80^{\circ}$C reduces but does not abolish long-term memory ($H=0.642\pm 0.016$), consistent with the earlier hysteresis results.\\

\section{Conclusions}

The aim of our study was to investigate the consequences of state-dependent channel affinity for different lipid species in demixing membranes. We found that when lipids are phase-separated, ion channels exhibit clustering, cooperativity, hysteresis, and long-term memory effects.

In particular, we have shown that fluctuations in local lipid composition give rise to effective attractive interactions and to correlations between ion channels even when they are far apart. These effects are responsible for the observed clustering tendency and cooperativity of activation. Moreover, since the typical relaxation times of these fluctuations is longer than any time-scale involved in channel activation, channel dynamics shows memory effects, including hysteresis of the activation curve.

The membrane influence provides explanations for previously puzzling experimental observations and generates testable predictions. In particular, fluorescence microscopy techniques could be used to measure the size distribution of clusters of ion channels for different lipid composition to correlate cluster properties to the phase behavior of the underlying membrane. Our model predicts that the largest clustering tendency occurs for near-critical membranes with a characteristic size distribution showing a power-law tail with a precise exponent. From the point of view of electrophysiology, we expect the slope of the G-V and Q-V curves (and their hysteresis effects) to show distinct dependencies on temperature and lipid composition, reflecting the fact that critical membranes enhance cooperativity in a predictable fashion. Similarly, the hypothesis of a long-range coupling mediated by lipids could be tested by measuring the influence of channel density on the slope of the activation curve. Finally, single channel recording experiments could be designed to characterize long-term memory in ionic currents for different temperatures and lipid compositions.  

Experimental confirmation of the paradigm proposed here would pave the way for a new class of effective models bridging membrane physics and physiology. Coupling ion channel conformational states and local lipid composition introduces a new level of complexity to the quantitative description of channel activity by taking into account the cellular context. For instance, the activity of several members of voltage-gated ion channels family is regulated by PIP2, a signaling lipid molecule that can facilitate or inhibit channel activation~\cite{dickson2019understanding} and that has been suggested to localize preferentially in lipid rafts~\cite{myeong2021compartmentalization}. The coupling between channel activation and rafts formation/disruption hypothesized here could have consequences on the PIP2 local concentration dynamics and thus potentially underlie some of the complex mechanisms employed by cells to modulate membrane excitability. Although the generality of these mechanisms is not yet fully understood, it is likely that state-dependent lipid coupling plays a significant physiological role, particularly in neurons where channels are abundant and interconnected~\cite{sato2019stochastic}. Specifically, clustering and cooperativity~\cite{dixon2022mechanisms} appear to have important functional consequences ~\cite{pfeiffer2020clusters}. Importantly, some of these mechanisms are likely at work in other types of membrane proteins with similar state-dependent affinity~\cite{levental2022regulation}, suggesting a universal class of self-organizing biomolecular systems. 

{\it Acknowledgements}

This research includes calculations carried out on Temple University's HPC resources and thus was supported in part by the National Science Foundation through major research instrumentation grant number 1625061 and by the US Army Research Laboratory under contract number W911NF-16-2-0189.

\bibliography{references.bib}

\newpage

% \textcolor{black}{Nota che nella c ci potrebbe essere una legge di potenza che scala come -2 come previsto da Machta}

% Intrinsic rates for each process are $\nu_L$ = 10 kHz (lipid translation), $\nu_C$ = 5 kHz (channel translation), and $\nu_R$ = 0.5 kHz (channel rotation).

\end{document}